# Phase-sensitive terahertz imaging using room-temperature near-field nanodetectors

Maria C. Giordano[*,1], Leonardo Viti[*,1], Oleg Mitrofanov[+,2] and Miriam S. Vitiello[#,1]

[1]*NEST, CNR - Istituto Nanoscienze and Scuola Normale Superiore, piazza San Silvestro 12, I-56127 Pisa (Italy)*
[2]*University College London, Electronic and Electrical Engineering, London, WC1E 7JE, UK*

[#]*Corresponding author*: miriam.vitiello@sns.it
[+]*Corresponding author*: o.mitrofanov@ucl.ac.uk

**\* authors contributed equally to this work**

**Imaging applications in the terahertz (THz) frequency range are severely restricted by diffraction. Near-field scanning probe microscopy is commonly employed to enable mapping of the THz electromagnetic fields with sub-wavelength spatial resolution, allowing intriguing scientific phenomena to be explored such as charge carrier dynamics in nanostructures and THz plasmon-polaritons in novel 2D materials and devices. High-resolution THz imaging, so far, has been relying predominantly on THz detection techniques that require either an ultrafast laser or a cryogenically-cooled THz detector. Here, we demonstrate coherent near-field imaging in the THz frequency range using a room-temperature nanodetector embedded in the aperture of a near-field probe, and an interferometric optical setup driven by a THz quantum cascade laser (QCL). By performing phase-sensitive imaging of strongly confined THz fields created by plasmonic focusing we demonstrate the potential of our novel architecture for high-sensitivity coherent THz imaging with sub-wavelength spatial resolution.**

Sub-wavelength resolution near-field imaging techniques in the infrared (IR) and terahertz (THz) ranges have recently shown an incredible potential in a variety of application fields ranging from fundamental light-matter interaction studies in nanostructures [1-5] to biological and chemical sciences [6-9], where high sensitivity combined with non-invasive subwavelength probing is required. The spectrum of THz near-field imaging applications is growing and it includes nanoscale mapping of plasmons in emerging bi-dimensional (2D) atomic materials (topological insulators [10], phosphorene [3], silicene [11], and their combined van der Waals heterostructures [12]), fundamental studies of plasmonic devices and coherent probing of sub-wavelength size ($<\lambda/10$) resonators [13-16]. This research feeds into engineering of novel THz optical components, such as negative refractive index materials, magnetic mirrors and filters [17-19].

To date, different near-field probing schemes have been developed and implemented for imaging systems [13-28], exploiting either scattering tip probes (known as apertureless probes) [23-28] for achieving nanometer-level resolution, or sub-wavelength size metallic aperture probes (a-SNOM) [14-16,22], electro-optic probes [18,20], and miniaturized photoconductive detectors [13,19]. The latter approaches are highly versatile and robust for large-scale (100 μm – 3 mm scale) near-field subwavelength resolution THz microscopy and spectroscopy, and they have enabled investigations of macroscopic THz devices (including metamaterials [17-19], waveguides [29], and resonators [13-16]) and inspection of biological tissues [6]. Coherent detection, which captures the intensity and phase information, proved essential for THz near-field microscopy. Most near-field mapping experiments at THz frequencies reported so far utilized the phase-sensitive THz time domain spectroscopy (TDS) techniques, which however require costly ultrafast lasers [30]. In the visible and near/mid-IR frequency ranges, alternative coherent architectures have been devised, including interferometric and holographic approaches [31]. Coherent detection in the latter cases is typically achieved by converting or scattering the evanescent waves from the near-field region into propagating waves, and then detecting them in the far-field using interferometers.

Here, we demonstrate phase-sensitive near-field THz imaging at room-temperature enabled by a simple and versatile interferometric optical setup with a THz QCL and an aperture type near-field probe. A nanowire (NW) thermoelectric detector is integrated into the aperture enabling subwavelength resolution and coherent gain for improved sensitivity. We thus eliminate the cost and complexity of the

TDS system by employing a THz QCL and a room-temperature detector, achieving phase-sensitive near-field imaging with subwavelength resolution. The imaging system architecture can be exploited with other coherent THz sources and a range of different nanoscale THz detectors. Additionally, the same interferometric setup enables spectral analysis of the THz field.

In aperture-type microscopy, spatial resolution is determined by the aperture size, $a$, however the possibility to achieve resolution smaller than 1/100 of the wavelength is practically limited by strong suppression of aperture transmission $T$, which follows the power law $T \sim a^6$ [32,33]. A promising approach for improving spatial resolution in a-SNOM schemes relies on the reduction of the detector size, and placing it in the proximity of the aperture [33]. We recently demonstrated that the dramatic drop in aperture transmission can be mitigated by integrating a THz thermoelectric nanodetector [34,35] inside the sub-wavelength aperture. Sensitivity of such a near-field probe rises due to detection of the evanescent components of the aperture transmitted wave [33,34].

To address the need for coherent THz imaging we designed a simple interferometric optical system shown in Fig. 1a. We exploited a QCL operating at 3.4 THz, which was driven in pulsed mode regime (2% duty cicle) to deliver only 28 µW of avarage power. The QCL beam was collimated with a Picarin lens (L) having focal length of 25 mm. The beam was then split into two, $I_1$ and $I_2$: one for illumination of the sample and the other for providing a reference for coherent detection. A nanoscale THz detector was embedded into a near-field probe, which was positioned in close proximity of the sample, as shown in the inset of Fig. 1a, with the reference beam illuminating the probe from its back side. The THz field of the sample then couples into the aperture of the probe, where it interferes with the reference field. The detector therefore senses the superposition of the local field collected by the probe aperture and the reference field. Phase difference between the two fields can be adjusted by an optical delay stage inserted into the reference beam path.

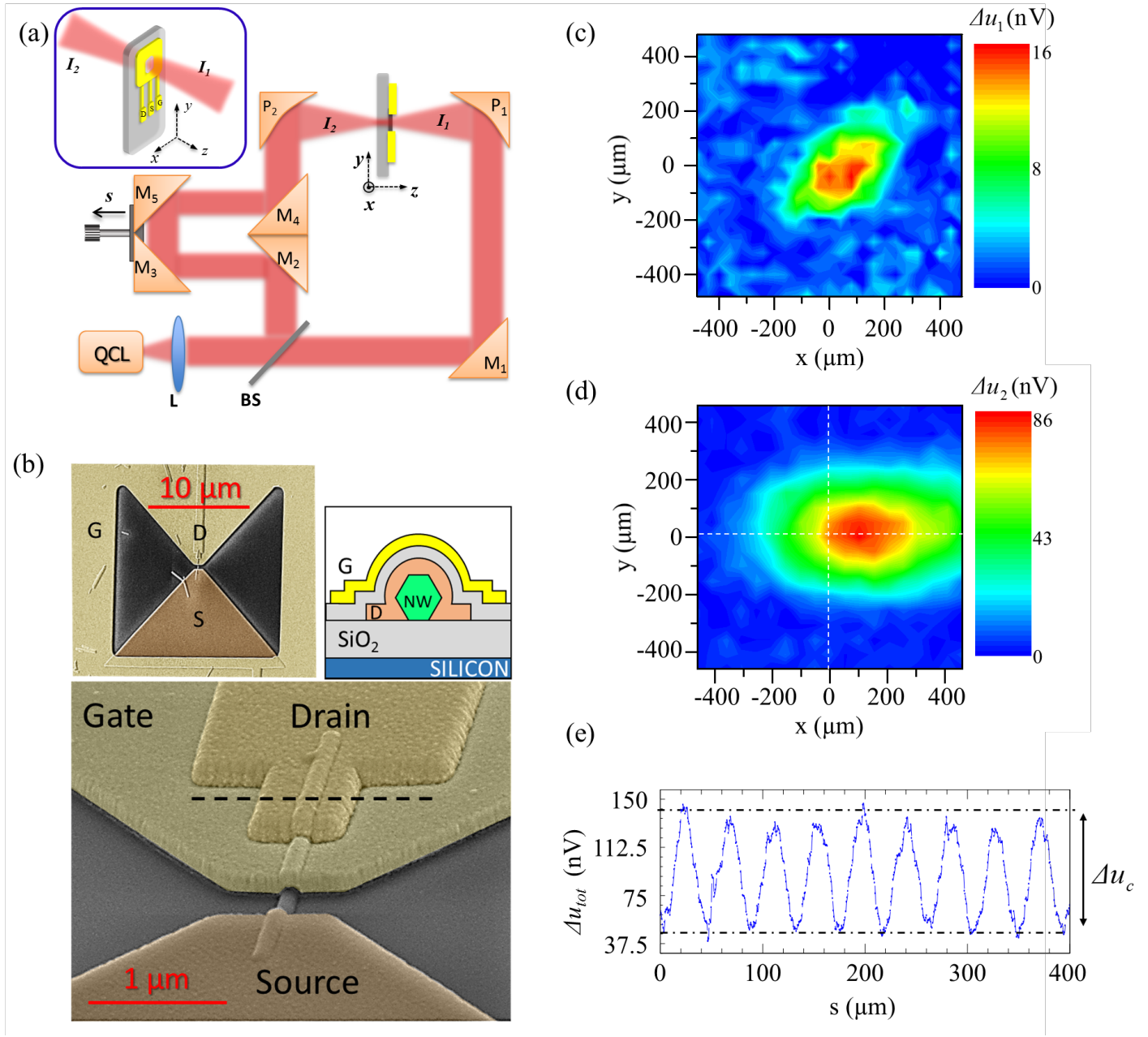

**Figure 1**: **(a)** Schematics of the interferometric THz near-field microscopy setup. The inset shows the the near-field probe geometry **(b)** (upper left and lower left) Scanning electron microscope (SEM) images of the near-field probe with an embedded FET-based THz nanodetector (view angles of 0° and 70°). A top gate contact (G) defines the aperture; the aperture size is 18 μm × 18 μm; the InAs nanowire detector is at the aperture center, the source (S) and drain (D) contacts are isolated from the gate with a layer of $SiO_2$; (upper right) schematic diagram of the cross-sectional view of the detector. **(c)** Spatial distribution of the detected photovoltage $\Delta u_1$ for the front ($I_1$) and **(d)** back side ($I_2$) illumination; **(e)** Interference trace acquired when the front and the reference beams simultaneously illuminate the near-field probe positioned in the focal plane at the center of the 2D scan area, and the relative phase is tuned using the delay stage.

In our experiment the collimated THz beam was directed towards an undoped Si wafer acting as a beam splitter (BS). The transmitted beam was then focused in the sample region with a parabolic mirror ($P_1$) having focal length of 25 mm. The reflected beam (reference beam) reached the optical delay stage, which consisted of two pairs of perpedincular flat mirrors: $M_2$, $M_4$ (fixed) and $M_3$, $M_5$ (movable). Translation of the delay stage ($\Delta s$) enabled tuning the optical path length, thus varying the relative phase between the reference beam and the illumination beam. A parabolic mirror ($P_2$) with the focal length of 50 mm focused the reference beam in the sample region from the opposite side (see Fig.1a).

In the sample region, the near-field probe was supported by an automated *xy*-translation stage, which enabled positioning and scanning of the probe in the focal plane, and by a *z*-axis translation stage for controlling the sample-probe separation. The near-field probe had a 18 μm input aperture and an InAs nanowire-based THz detector embedded inside the aperture (Fig. 1b), as described in Ref. [34]. The detector is sensitive to the THz waves incident from both sides of the aperture [34] and therefore it allows us to obtain a coherent superposition of the sample and the reference field, as well as the intensity maps of the sample field ($I_1$) and the reference beam ($I_2$).

The THz detector, in our experiment, was a field-effect transistor (FET), fabricated by nano-lithography, with an epitaxially grown InAs NW as the active channel (see Supplementary Material). The NW was integrated within the aperture area with the source and drain contacts connected to its ends. The side of the NW opposite of the source contact was covered by a trapezoid-shaped extension from the aperture edge, serving as the transistor gate (Fig. 1b). This geometry asymmetrically feeds the incident THz radiation into the FET channel and induces a temperature gradient $\Delta T$ along the NW [34]. The break in the symmetry enables thermoelectric THz detection with the possibility to enhance the device responsivity by applying a gate voltage [34]. A steady-state thermoelectric voltage $\Delta u = S_b \Delta T$ (where $S_b$ is the Seebeck coefficient) arises across the channel between the source and drain electrodes.

This voltage is proportional to the intensity of the coherent superposition of the THz field coupled through the aperture and the reference field.

Figures 1c and 1d show profiles of the THz beam incident on the aperture ($I_1$) and the reference beam incident on the back side of the detector ($I_2$). The profiles of $I_1$ and $I_2$ had elliptical shapes in the focal plane. The distribution of $I_1$ showed the incident beam close to the center of the scanned area. The distribution of the reference beam, $I_2$, was wider and it was slightly displaced (approximately 100 µm) with respect to the center. The beam size, defined as the full width at half maximum (FWHM) of the optical signal retrieved along the x and y axes, was about 400 µm × 260 µm for the incident front beam ($I_1$), and 520 µm × 280 µm for the reference beam ($I_2$). The two beams interfered within the area of the overlap. The total power in the incident front beam and in the reference beam, measured with a pyroelectric detector, were $P_{tot(1)}$= 300 nW and $P_{tot(2)}$=600 nW, respectively. We then determined the responsivity according to the equation:

$$R_{(i)} = \frac{\Delta u_{(i)}}{P_{tot(i)}} \times \frac{S_{tot(i)}}{S_a} \quad \text{with } i=1,2 \tag{1}$$

Here the letter $i$ distinguishes between the front ($i = 1$) and the reference ($i = 2$) beam. $S_{tot(i)}$ is the beam spot area, approximated as the area of an ellipse having the two axes corresponding to the radii of the gaussian profile $r_{(i)} = \frac{FWHM_{(i)}}{2.355}$ of the intensity maps shown in Figs. 1c,d and $S_a$ is the aperture area [34]. The detected optical intensity maps (Figs. 1 c,d) thus allow evaluation of the maximum responsivities $R_1$=9.7 V/W and $R_2$=36.5 V/W on the aperture side and on the back side, respectively. The corresponding maximum signals were $\Delta u_1$ = 16 nV and $\Delta u_2$ = 86 nV.

The interference between the two beams produced a superposition photovoltage signal $\Delta u_{tot}$, which varied in the range of 50 nV - 140 nV as a function of the optical path difference ($s$) (Fig. 1e), following a sinusoidal function with periodicity $p \sim 43.5$ µm, which equals to one half of the QCL wavelength ($\lambda_{QCL}/2$). The maxima and minima correspond to the constructive and destructive interference between the two beams illuminating the near-field probe on the aperture and on the reference sides, respectively. Since the periodic behavior was stable over a phase delay shift exceeding 1 mm (Supplementary Material, Fig. S1a), the detected interferogram was additionally exploited to extract the Fourier spectrum of the THz QCL source (Supplementary Material, Fig. S1 b,c). This represents a further functionality of this simple interferometric setup that can find applications in QCLs characterization.

The periodic behavior also clearly demonstrates the coherent superposition of the two beams locally detected by the near-field probe. Indeed, the total intensity can be written as the superposition of the two beams ($I_1$ and $I_2$) as:

$$I_{tot} = I_1 + I_2 + 2\gamma\sqrt{I_1 I_2}\cos(\delta\varphi) \qquad (2)$$

where the last term accounts for the coherent optical interference between the two beams with a phase difference $\delta\varphi$, set by the delay stage, and the parameter $\gamma$ describes the degree of coherence of the two beams in the range $0 \leq \gamma \leq 1$. The detected photovoltage signal $\Delta u_{tot}$ is related to the optical intensity $I_1$ and $I_2$ coupled to the device [34] as:

$$\Delta u_{tot} = R_1 I_1 + R_2 I_2 + 2\gamma\sqrt{R_1 I_1 R_2 I_2}\cos(\delta\varphi) \qquad (3)$$

From this equation the oscillation amplitude of the interference pattern (Fig. 1e) can be defined as a coherent photovoltage, $\Delta u_c = \Delta u_{tot}^{in-phase} - \Delta u_{tot}^{out-of-phase} = 90$ nV, which corresponds to the coherent term of the total detected photovoltage $\Delta u_c = 4\gamma\sqrt{R_1 I_1 R_2 I_2}$, and is proportional to the coherent component of the optical intensity.

Therefore the coherent signal $\Delta u_c$ carries the information about the amplitude of the electric field proportional to $\sqrt{I_1}$ and the phase relative to the reference beam $\delta\varphi$. To make the coherent signal independent of the spatial variation of the reference beam, the intensity distribution $I_2$ and the phase $\varphi$ need to remain constant within the image area.

To demonstrate phase sensitive THz imaging with spatial resolution beyond the diffraction limit, we selected an object, which produces both a phase variation across the image area and a sub-wavelength size amplitude variation. Such a field distribution can be formed by focusing the THz beam into a deeply sub-wavelength volume with surface plasmon guiding along two metallic needles [36,37]. We employed two PtIr needles having a diameter of 500 μm and mechanically polished to obtain a tip apex radius smaller than 1 μm. The needles were fixed at a relative angle of ~ 60° and at a relative distance between their tips of ~10 μm [37], by means of a micrometer-controlled stage. This twin-needle configuration was oriented in the *yz*-plane and positioned in front of the aperture (Fig. 2a).

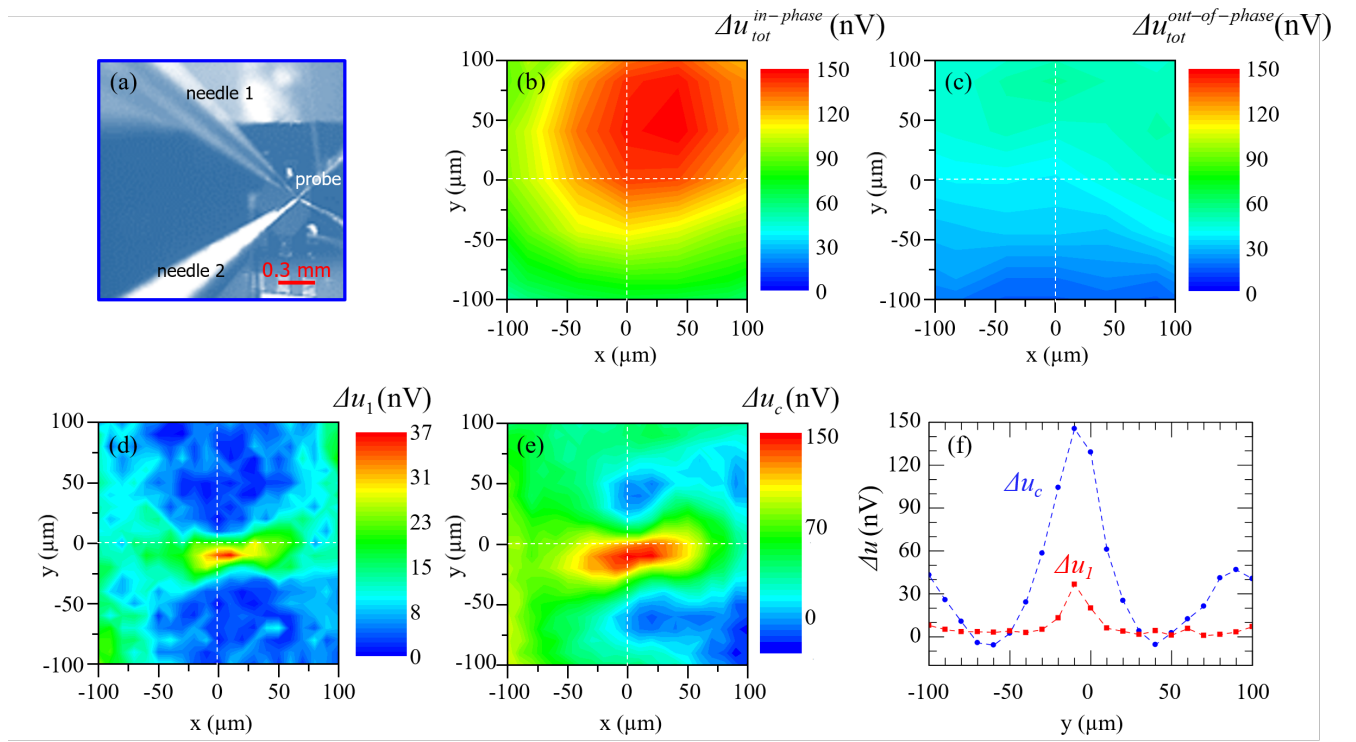

**Figure 2**: **(a)** Optical image of two needles employed for focusing the THz beam to a subwavelength spot. The needles are placed in front of the NW nano-detector probe. **(b,c)** Spatial distribution of the near-field probe photovoltage $\Delta u_{tot}$, for the in-phase and out-of-phase conditions without the needles. **(d)** Spatial distribution of $\Delta u_1$, detected under the front side only ($I_1$) illumination of the near-field probe, and **(e)** spatial distribution of the coherent photovoltage $\Delta u_c = \Delta u_{tot}^{in-phase} - \Delta u_{tot}^{out-of-phase}$ when the two needles focus the incident beam to a sub-wavelength spot. **(f)** Vertical line scans of $\Delta u_1$ and $\Delta u_c$ extracted from maps (d) (red line) and (e) (blue line), respectively.

We first verified that the interference of the incident and the reference beams did not produce strong variations of phase and amplitude within the image area. Figures 2b and 2c show the zoomed 2D profiles of the detected photovoltage $\Delta u_{tot}$, generated by the two waves for two different positions of the delay stage $s$, corresponding to a maximum and a minimum of the interference pattern (Fig.1e) measured at the image center. These delay stage positions will be referred to the in-phase and out-of-phase conditions, respectively. The in-phase interference pattern, $\Delta u_{tot}^{in-phase}$ (Fig. 2b), shows a spot of about 200 μm diameter with a maximum (of 150 nV) close to the center of the scan. The out-of-phase interference intensity map, $\Delta u_{tot}^{out-of-phase}$ (Fig. 2c), is characterized by an average signal of ~ 40 nV slowly varying by about ±10 nV across the image. Although the maps of $\Delta u_{tot}^{in-phase}$ and $\Delta u_{tot}^{out-of-phase}$ are not uniform across the 200 μm × 200 μm image area, their variation is small on the scale of the confined electric field focused by the two needles. The extended 2D maps of $\Delta u_{tot}$ are shown in Supplementary Materials, Fig. S1 d,e.

A set of 2D scans was then acquired under the illumination of the probe on its front side only ($I_1$), while the distance between the tips and the near-field probe aperture was gradually decreased. This allowed us to achieve localization of the THz intensity when the tip-aperture distance was reduced below 10 μm (data shown in Supplementary Material, Fig. S2 a-d). The photovoltage map obtained under the $I_1$ illumination is shown in Fig. 2d. A strong THz field localization, created by guiding THz waves along the needles, was detected in proximity of the center of the scan. The needles increased the maximum photovoltage by about +60%. The sub-wavelegth localization [37] allowed us to evaluate the spatial resolution of the aperture probe defined as the FWHM along the y scan direction. The FWHM (~ 17 μm) is comparable with the aperture size, confirming the capability of the NW probe to resolve sub-wavelength size features.

Figure 2e shows the coherent signal photovoltage $\Delta u_c = \Delta u_{tot}^{in-phase} - \Delta u_{tot}^{out-of-phase}$ detected when $I_1$ and $I_2$ simultaneously illuminate the near-field probe. The detected $\Delta u_c$ signal is confined along the *y* axis in proximity of the center of the scanned area, in analogy to the case of Fig. 2d, however it shows a broader maximum with respect to $\Delta u_1$. This apparent discrepancy is inherent to the methodology that extracts the electric field distribution, instead of the electromagnetic intensity distribution. We note that the aperture size limits the spatial resolution of these images to ~17 μm. Additionally, the coherent image shows two minima above and below the central peak. These minima do not appear in the incoherent image (Fig.2d). From this measurement it is possible to retrieve the phase image of an object placed in front of the aperture (Supplementary Material, Fig.S4d). Furthermore the amplitude of the detected signal is approximately four times higher compared to the signal $\Delta u_1$. We highlight the difference in Fig. 2f, where line scans along the vertical axis are shown for $\Delta u_1$ and $\Delta u_c$.

In order to interpret the coherent image we refer to Eq. 3, which shows that the coherent signal $\Delta u_c$ is proportional to the electric field amplitude $\sqrt{I_1} \propto E_1$, and depends on the relative phase difference as $\cos(\delta\varphi)$. The image in Fig. 2d, on the other hand, provides only the time-averaged intensity of the field. The field pattern in Fig. 2e represent the launching of surface plasmon waves, which travel within the image plane away from the needle tips. This leads to the variation of the coherent signal along the *y*-axis [37]. The distance between the central maximum (the launching point) and the minima is approximately 50 μm corresponding to ~ $\lambda_{QCL}/2$. In the intensity image, on the other hand, we expect to see a monotonic decay of intensity away from the central point, however the noise level was too high to detect it (Fig. 2e). It must be noted that the signal $\Delta u_c = \Delta u_{tot}^{in-phase} - \Delta u_{tot}^{out-of-phase}$ removes the incoherent component of the detected signal. The coherent image thus provides additional phase information.

In addition to proving the phase information, the coherent detection improves the sensitivity, thus allowing us to achieve a wider dynamic range. In order to evaluate the sensitivity limit, here we introduce a framework for describing the detector responsivity. As it follows from Eq. 2, the measured intensity $I_{tot}$ depends on the relative phase $\delta\varphi$. We therefore introduce effective responsivity $R^*$ defined as:

$$R^* = \frac{R_1 I_1 + R_2 I_2 + 2\gamma\sqrt{R_1 I_1 R_2 I_2}\cos(\delta\varphi)}{I_1 + I_2} \qquad (4)$$

And the phase-dependent component $R_c$:

$$R_c = \frac{4\gamma\sqrt{R_1 I_1 R_2 I_2}}{I_1 + I_2}, \qquad (5)$$

which will be referred to as the coherent componenent of effective responsivity. To better clarify Eq. 4 and 5, we plot in Fig. 3a the ratio $R^*/R_2$ as a function of the dimensionless parameter $\xi = (I_2-I_1)/(I_2+I_1)$, which accounts for the asymmetric illumination of the probe on its front and back sides. $R^*(\xi)$ in general follows an ellipse function, which depends on the degree of coherence of the two beams and on the ratio of responsivities $R_1$ and $R_2$. For any combination of $I_1$ and $I_2$, the effective responsivity $R^*$ varies with the relative phase $\delta\varphi$ between the two limits defined by points of intersection of the ellipse with a vertical line $\xi = (I_2-I_1)/(I_2+I_1)$. This span corresponds to the coherent responsivity $R_c$.

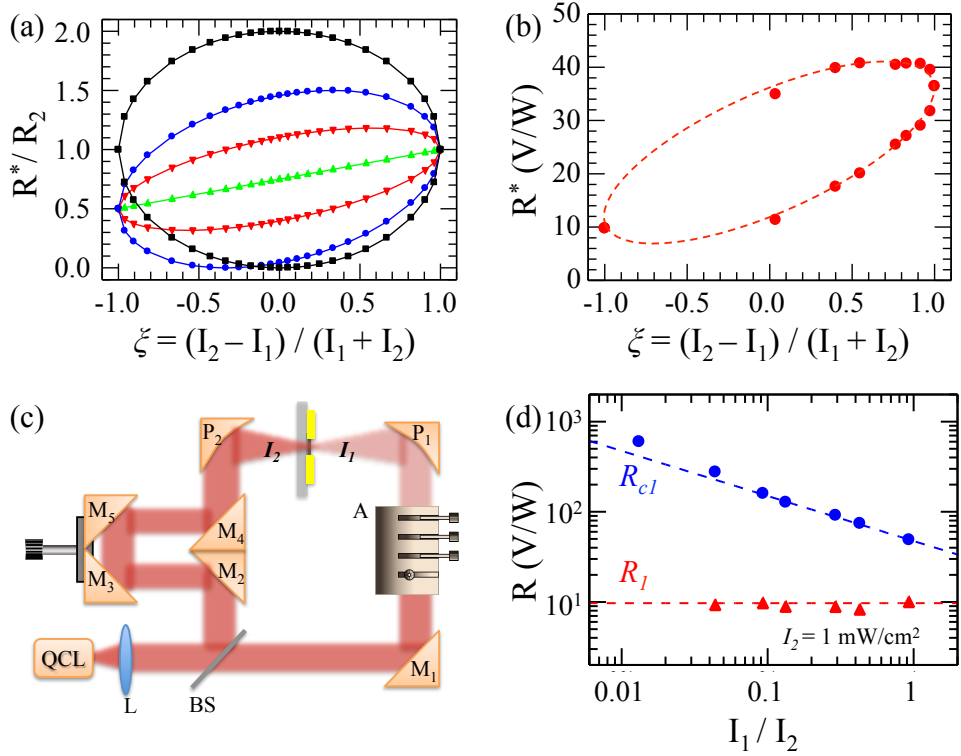

**Figure 3**: **(a)** Graphic representation of the responsivity ratio $R^*/R_2$ as a function of $\xi$ for different experimental configurations. **(b)** Measured responsivity $R^*$ as a function of the degree of asymmetry of the illuminatoin $\xi$. **(c)** Schematics of the interferometric optical setup with an additional attenuator (A) positioned along the optical path of the front illumination beam ($I_1$). **(d)** Coherent responsivity ($R_{c1}$, blue circles) and non-coherent responsivity ($R_1$, red triangles) plotted as a function of the intensity ratio $I_1/I_2$.

In the case of the coherent beam ($\gamma = 1$) and symmetric device responsivity, $R_1 = R_2$, the ellipse is symmetric with respect to the $\xi = 0$ axis (black line in Fig. 3a). At the two extrema $\xi = 1$ ($I_1 = 0$) and $\xi = -1$ ($I_2 = 0$) the effective responsivity becomes the incoherent responsivity $R^* = R_2$ and $R^* = R_1$, respectively. For $\xi = 0$, the interference between the two beams can be controlled by moving the delay stage and $R^*$ can be continuously varied between zero and $2R_2$.

If the device architecture is asymmetric ($R_1 \neq R_2$), as in the case of the near field probe, the ellipse is tilted in the $\xi$-$R^*$ plane. For example, the blue curve in Fig. 3a illustrates the case $R_1 = R_2/2$. For partially coherent beams, when $\gamma < 1$, the ellipse shrinks along the vertical axis (red curve in Fig. 3a). In the extreme case of $\gamma = 0$ the ellipse collapses into a line connecting the two points ($\xi = -1$, $R^* =$

$R_1$) and ($\xi =1$, $R^*= R_2$) (green line in Fig. 3a). Under this condition the detected signal is independent of the delay stage position because the phases of the front and reference waves are uncorrelated, and therefore the detector signal is proportional to the sum $I_1+I_2$.

We measured the degree of coherence of the front and reference beams by adjusting the optical intensity $I_1$ with THz attenuation filters while keeping the reference intensity $I_2$ constant (Fig. 3c). In this way we varied the degree of asymmetry of the probe illumination, $\xi$. In particular, we measured the effective responsivity of the probe, $R^*$, by changing $I_1$ between 12.9 µW·cm$^{-2}$ and 922 µW·cm$^{-2}$ ($P_{tot(1)}$ between 7.6 nW and 543 nW), while $I_2$ was kept constant at 989 µW·cm$^{-2}$ ($P_{tot(2)}$ = 816 nW). A constellation of the measured points follows an elliptical shape (Fig. 3b) with eccentricity $\gamma = 0.63$, corresponding to the degree of coherence of the two beams.

Using this framework, we can quantitatively evaluate the advantage of coherent detection by means of estimating the effective responsivity of the near-field probe when $I_1$ is only a small fraction of $I_2$. This situation well represents a practical near-field imaging system in which the field coupled through the aperture can be weak. We can then define the coherent responsivity ($R_{c1}$) of the probe to the intensity coupled into the aperture $I_1$ as:

$$R_{c1} = \frac{\Delta u_c}{I_1} = R_c \frac{I_1+I_2}{I_1} = 4\gamma \sqrt{R_1 R_2 \frac{I_2}{I_1}} \qquad (6)$$

This quantity depends on the ratio $I_1/I_2$ and it decreases when $I_1$ is raised for a fixed value of $I_2$. The measured values of $R_1$ and $R_{c1}$ are plotted in Fig. 3d as functions of the ratio $I_1/I_2$ and clearly show the advantage of the coherent detection scheme with respect to the standard front illumination approach. Indeed, under the front illumination only, the minimum detected intensity is $I_1 = 40$ µW·cm$^{-2}$. This sets an upper limit for the minimum detectable power (MDP) of 25.3 nW. At $I_1 = 12.9$ µW·cm$^{-2}$, no THz signal is detected in the standard detection scheme, i.e. the THz power is below the MDP. On the other hand, if a coherent detection scheme is employed, a power law behavior for $R_{c1}$ was observed:

$R_{c1} \propto \sqrt{\frac{I_2}{I_1}}$. The minimum detected intensity was reduced to 12.9 µW·cm$^{-2}$, corresponding to a coherent photovoltage $\Delta u_c$ of 26 nV. This gives an upper limit for the MDP of 7.6 nW, clearly showing that the coherent detection scheme effectively lowers the MDP level and strongly improves the sensitivity of near-field measurements.

In conclusion, we demonstrated interferometric THz near-field imaging with sub-wavelength spatial resolution using a THz *a*-SNOM probe with a monolithically embedded coherent detector. The detector is based on an InAs NW and it is integrated into a 18 µm aperture. We showed the advantages of the interferometric detection scheme to perform sub-wavelength resolution amplitude and phase imaging with a MDP lower than 7.6 nW. Our results pave the way to development of new coherent THz microscopes for large-area sub-wavelength resolution phase- and amplitude-sensitive imaging. In combination with QCL sources operating in the 1.5–5.0 THz range, this imaging technique can aid in the development of novel THz components (mirrors, filters, metamaterials, metalenses and sub-wavelength resonators) and open new research avenues in the studies of fundamental light-matter interaction phenomena in many interdisciplinary fields crossing optics, photonics, chemistry and biology.

**Acknowledgements**

The authors acknowledge support from the European Union via the ERC Grant SPRINT, the Far-FAS program DIAST and the Royal Society and the EPSRC. M.S.V. acknowledge financial support from the second half of the Balzan Prize 2016 in applied photonics delivered to Federico Capasso. The authors acknowledge D. Ercolani and L. Sorba for the growth of InAs nanowires.

See Supplement 1 for supporting content.

# References

1. Z. Fei, A.S. Rodin, G. O. Andreev, W. Bao, A. S. McLeod, M. Wagner, L. M. Zhang, Z. Zhao, M. Thiemens, G. Dominguez, M. M. Fogler, A. H. Castro Neto, C. N. Lau, F. Keilmann, and D. N. Basov, "Gate-tuning of graphene plasmons revealed by infrared nano-imaging," Nature **487**, 7405 (2012).

2. R. Jacob, S. Winnerl, M. Fehrenbacher, J. Bhattacharyya, H. Schneider, M. T. Wenzel, H. G. von Ribbeck, L. M. Eng, P. Atkinson, O. G. Schmidt, and M. Helm, "Intersublevel Spectroscopy on Single InAs-Quantum Dots by Terahertz Near-Field Microscopy," Nano Lett. **12**, 4336 (2012).

3. M. A. Huber, F. Mooshammer, M. Plankl, L. Viti, F. Sandner, L. Z. Kastner, T. Frank, J. Fabian, M. S. Vitiello, T. L. Cocker and R. Huber, "Femtosecond photo-switching of interface polaritons in black phosphorus heterostructures," Nat. Nanotech. **12**, 3 (2017).

4. O. Schubert, M. Hohenleutner, F. Langer, B. Urbanek, C. Lange, U. Huttner, D. Golde, T. Meier, M. Kira, S. W. Koch and R. Huber, "Sub-cycle control of terahertz high-harmonic generation by dynamical Bloch oscillations," Nat. Photon. **8**, 2 (2014).

5. J. Vedran, K. Iwaszczuk, P. H. Nguyen, C. Rathje, G. J. Hornig, H. M. Sharum, J. R. Hoffman, M. R. Freeman, and F. A. Hegmann, "Ultrafast terahertz control of extreme tunnel currents through single atoms on a silicon surface," Nat. Phys. **13**, 591 (2017).

6. H. Chen, W.-J. Lee, H.-Y Huang, C.-M. Chiu, Y.-F.Tsai, T.-F.Tseng, J.-T. Lu, W.-L. Lai and C.-K. Sun, "Performance of THz fiber-scanning near-field microscopy to diagnose breast tumors," Optics Express **19**, 19523 (2011).

7. M.-A. Brun, F. Formanek, A. Yasuda, M. Sekine, N. Ando, and Y. Eishii, "Terahertz imaging applied to cancer diagnosis," Phys. in Med. and Biol. **55**, 16 (2010).

8. L. Rong, T. Latychevskaia, C. Chen, D. Wang, Z. Yu, X. Zhou, Z. Li, H. Huang, Y. Wang, and Z. Zhou, "Terahertz in-line digital holography of human hepatocellular carcinoma tissue," Sci. Rep. **5**, 8445 (2015).

9. L.V. Titova, A. K. Ayesheshim, A. Golubov, R. Rodriguez-Juarez, R. Woycicki, F. A. Hegmann, and O. Kovalchuk, "Intense THz pulses down-regulate genes associated with skin cancer and psoriasis: a new therapeutic avenue?," Sci. Rep. **3**, 2363 (2013).

10. M. Autore, H. Engelkamp, F. D'Apuzzo, A. Di Gaspare, P. Di Pietro, I. Lo Vecchio, M. Brahlek, N. Koirala, S. Oh, and S. Lupi, "Observation of Magnetoplasmons in $Bi_2Se_3$ Topological Insulator," ACS Photonics **2**, 1231-1235 (2015).

11. L. Tao, E. Cinquanta, D. Chiappe, C. Grazianetti, M. Fanciulli, M. Dubey, A. Molle and D. Akinwande, "Silicene field-effect transistors operating at room temperature," Nat. Nanotech. **10**, 227–231 (2015).

12. T. Low, A. Chaves, J. D. Caldwell, A. Kumar, N. X. Fang, P. Avouris, T. F. Heinz, F. Guinea, L. Martin-Moreno and F. Koppens, "Polaritons in layered two-dimenisonal materials," Nat. Mater. **16**, 182-194 (2017).

13. A. Bhattacharya and J. Gómez Rivas, "Full vectorial mapping of the complex electric near-fields of THz resonators," APL Photonics **1**, 8 (2016).

14. O. Mitrofanov, I. Khromova, T. Siday, R. Thompson, A. Ponomarev, I. Brener, and J. L. Reno, "Near-Field Spectroscopy and Imaging of Subwavelength Plasmonic Terahertz Resonators," IEEE Trans. Terahertz Sci. Technol. **6**, 382 (2016).

15. O. Mitrofanov, Z. Han, F. Ding, S. I. Bozhevolnyi, I. Brener, and J. L. Reno, "Detection of internal fields in double-metal terahertz resonators," Appl. Phys. Lett. **110**, 061109 (2017).

16. I. Khromova, P. Kužel, I. Brener, J. L. Reno, U.-C. Chung Seu, C. Elissalde, M. Maglione, P. Mounaix, and O. Mitrofanov, "Splitting of magnetic dipole modes in anisotropic TiO2 micro-spheres," Laser Photonics Rev. **10**, 681 (2016).

17. S. H. Lee, J. Choi, H.-D. Kim, H. Choi and B. Min, "Ultrafast refractive index control of a terahertz graphene metamaterial," Scientific Reports **3**, 2135 (2013).

18. N. Kumar, A. C. Strikwerda, K. Fan, X. Zhang, R. D. Averitt, P. C. M. Planken, and A. J. L. Adam, "THz near-field Faraday imaging in hybrid metamaterials," Opt. Express **20**, 11277 (2012).

19. A. Bitzer, A. Ortner, H. Merbold, T. Feurer, and M. Walther, "Terahertz near-field microscopy of complementary planar metamaterials: Babinet's principle,' Opt. Express 19, 2537 (2011).

20. F. Blanchard, K. Ooi, T. Tanaka, A. Doi, and K. Tanaka, "Terahertz spectroscopy of the reactive and radiative near-field zones of split ring resonator," Opt. Express 20, 19395 (2012).

21. R.I. Stantchev, D.B. Phillips, P. Hobson, S.M. Hornett, M.J. Padgett, and E. Hendry, "Compressed sensing with near-field THz radiation," Optica 4, 989-992 (2017);

22. E. Ash and G. Nicholls, "Super-resolution aperture scanning microscope," Nature **237**, 510 (1972).

23. J. Wessel "Surface-enhanced optical microscopy." JOSA B **2**, 1538-1541 (1985).

24. F. Keilmann and R. Hillenbrand, "Near-field microscopy by elastic light scattering from a tip," Phil. Trans. R. Soc. Lond. A **362**, 787-805 (2004).

25. T. Taubner, D. Korobkin, Y. Urzhumov, G. Shvets and R. Hillenbrand, "Near-field microscopy through a SiC superlens," Science **313**, 5793, (2006).

26. M. Eisele, T. L. Cocker, M. A. Huber, M. Plankl, L. Viti, D. Ercolani, L. Sorba, M. S. Vitiello and R. Huber, "Ultrafast multi-terahertz nano-spectroscopy with sub-cycle temporal resolution," Nat. Photon. **8**, 841–845 (2014).

27. P. Dean, O. Mitrofanov, J. Keeley, I. Kundu, L. Li, E. H Linfield, A. G. Davies "Apertureless near-field terahertz imaging using the self-mixing effect in a quantum cascade laser," Appl. Phys. Lett. 108, 091113 (2016).

28. R. Degl'Innocenti, R. Wallis, B. Wei, L. Xiao, S. J. Kindness, O. Mitrofanov, P. Braeuninger-Weimer, S. Hofmann, H. E. Beere, D. A. Ritchie, "Terahertz Nanoscopy of Plasmonic Resonances with a Quantum Cascade Laser," ACS Photonics 4:9, 2150-2157 (2017).

29. O. Mitrofanov, T. Tan, P. R. Mark, B. Bowden, and J. A. Harrington, "Waveguide mode imaging and dispersion analysis with terahertz near-field microscopy," Appl. Phys. Lett. **94**, 171104 (2009).

30. S. S. Dhillon, M. S. Vitiello, E. H. Linfield, A. G. Davies, M. C. Hoffmann et *al*., "The 2017 terahertz science and technology roadmap," J. Phys. D: Appl. Phys. **50**, 043001 (2017).

31. M. Schnell, P. S. Carney and R. Hillenbrand, "Synthetic optical holography for rapid nanoimaging," Nat. Comm. **5**, 3499 (2014).

32. H. Bethe "Theory of diffraction by small holes" Phys. Rev. **66**, 163 (1944).

33. O. Mitrofanov, I. Brener, T. S. Luk, and J. L. Reno, "Photoconductive Terahertz Near-Field Detector with a Hybrid Nanoantenna Array Cavity," ACS Photon. **2**, 1763-1768 (2015).

34. O. Mitrofanov, L. Viti, E. Dardanis, M. C. Giordano, D. Ercolani, A. Politano, L. Sorba and M. S. Vitiello, "Near-field terahertz probes with room-temperature nanodetectors for subwavelength resolution imaging," Sci. Rep. **7**, 44240 (2017).

35. L. Viti, J. Hu, D. Coquillat, W. Knap, A. Tredicucci, A. Politano and M. S. Vitiello, "Black Phosphorus Terahertz Photodetectors," Adv. Mater. **27**, 5567-5572 (2015).

36. M. Schnell, P. Alonso-Gonzalez, L. Arzubiaga, F. Casanova, L. E. Hueso, A. Chuvilin and R. Hillenbrand, "Nanofocusing of mid-infrared energy with tapered transmission lines," Nat. Photon. **5**, 283-287 (2011).

37. O. Mitrofanov, C. C. Renaud and A. J. Seeds, "Terahertz probe for spectroscopy of sub-wavelength objects," Optics Express **20**, 6197-6202 (2012).